\documentclass[journal]{rmaa}

\usepackage{psfrag,color}
\usepackage{graphicx}
\usepackage[latin1]{inputenc}

\title{A virtual observatory for photoionized nebulae: the Mexican Million Models database (3MdB).} 

\author{
  C. Morisset,\altaffilmark{1} 
  G. Delgado-Inglada,\altaffilmark{1}
  and N. Flores-Fajardo\altaffilmark{2}}

\altaffiltext{1}{Instituto de Astronomía, Universidad Nacional Autónoma de México}

\altaffiltext{2}{Department of Astronomy, Kavli Institute for Astronomy and Astrophysics, Peking University}

\shortauthor{Morisset, C., Delgado-Inglada, G., Flores-Fajardo, N.}
\shorttitle{3MdB: a virtual observatory for photoionized nebulae.}

\fulladdresses{
\item  Instituto de Astronomía, Universidad Nacional Autónoma de México, Apdo. Postal 70264, Méx. D. F., 04510 México
\item Department of Astronomy, Peking University, Beijing 100871, China}

\listofauthors{C. Morisset, G. Delgado-Inglada, N. Flores-Fajardo}
\indexauthor{Morisset, C.}
\indexauthor{Delgado-Inglada, G.}
\indexauthor{Flores-Fajardo, N.}

\abstract{Photoionization models obtained with numerical codes are widely used to study the physics of the interstellar medium (Planetary Nebulae, \ion{H}{ii} regions, etc). Grid of models are performed to understand what are the effects of the different parameters used to describe the regions on the observables (mainly emission line intensities). Most of the time, only a small part of the computed results of such grids are published, and they are sometimes hard to obtain in a user-friendly format. We present here the Mexican Million Models dataBase (3MdB), an effort of resolving both of these issues in the form of a database of photoionization models, easily accessible throught the MySQL protocol, and containing a lot of usefull outputs from the models, such as the intensities of 178 emission lines, the ionic fractions of all the ions, etc. Some examples of the use of the 3MdB are also presented. 
}

\resumen{Los modelos de fotoionizaci\'on obtenidos mediante c\'odigos num\'ericos son ampliamente utilizados para estudiar la f\'isica del medio interestelar (Nebulosas Planetarias, regiones \ion{H}{ii}, etc). Las redes de modelos de fotoionizaci\'on se calculan para entender cu\'ales son los efectos que tienen sobre los par\'ametros observables (principalmente intensidades de l\'ineas) los distintos par\'ametros que se utilizan para describir las nebulosas. Con frecuencia, s\'olo se publica una parte de los resultados calculados con estas redes de modelos y, a veces, es dif\'icil obtenerlos en un formato amigable. Aqu\'i presentamos la base de datos mexicana del mill\'on de modelos (3MdB), un esfuerzo para resolver ambos problemas en forma de una base de datos de modelos de fotoionizaci\'on, f\'acilmente accesible a trav\'es de un protocolo MySQL, y que contiene muchos par\'ametros de salida \'utiles como las 178 intensidades de l\'ineas de emisi\'on, las fracciones i\'onicas de todos los iones, etc. Adem\'as presentamos algunos ejemplos pr\'acticos de c\'omo usar 3MdB.
}

\addkeyword{H~II regions, Planetary Nebulae, Astronomical data bases, Galaxies: ISM }

\begin{document}
% Typeset article head
\maketitle

\section{Introduction}
\label{sec:intro}

One of the problems modern astronomers have to deal with when confronted to the managing of astronomical data is the sustainability of the data themselves.  This affects computational data as much as observational data.  This problem appears in two forms:  first, the format of the hardware medium can be obsolete: some research institutes keep maintained a very old computer out-of-date in terms of hardware and software just because it is the only one able to read some old tape formats, 3.5'' floppies, etc. In other places there is no way to access the data stored on those type of media.  Second, the format of the data in the file once read can be obsolete, like proprietary binary data formats that one cannot access anymore because the software used to write them is not available anymore.

 Concerning nebular astrophysics and its applications, many works rely on the consideration of grids of photoionization models. e.g.: \citet{1980Stasinska_aap84, 1996Stasinska_apjs107, 1997Korista_apjs108, 2001Charlot_mnra323,  2000Dopita_apj542, 2001Kewley_apj556, 2002Kewley_apjs142,  2004Groves_apjs153, 2006Dopita_apjs167, 2006Stasinska_mnra371, 2010Levesque_aj139, 2013Dopita_apjs208, 2014Perez-Montero_mnra441}
to quote just a few. In all these papers, the results of the computational effort have drastically been reduced into a few tables containing  only a tiny proportion  of the model results. 
This is a pity because it prevents the use of the computed models for other purposes not necessarily foreseen by the authors. In other cases, it is almost impossible to obtain the set of results from the published papers, only points in figures are available and there is no public access to the digital values. It is not always easy for astronomers to develop tools allowing to give access to their data by the community, although  this task is made much easier with the efforts of the CDS\footnote{Created in 1972 as the Stellar Data Centre and changed its name to Strasbourg astronomical Data Centre in 1983. \url{http://cds.u-strasbg.fr/}}. 
Statistics on the way astronomers share their data can be found in \citet{2014Goodman_PLoS10, 2014Pepe_PLoS9}. New tools have been developed for  data mining in  various areas of the Virtual Observatories\footnote{\url{http://www.ivoa.net/}}, but we have to admit that the community of the interstellar medium astrophysics is not very involved in the development and use of these tools.

The main purpose of the Mexican Million Models dataBase (3MdB) is to offer to the comunity (which includes modelers and observers) a free access to a huge amount of results from photoionization models obtained by running the {\sc CLOUDY} program \citep{2013Ferland_rmxa49}. The database is stored in an efficient way using a MySQL system. The disk space usage is then optimized and the requests are efficient, compared to what would be the result of a dedicated program looking into files. The format of the result transmitted to the user is versatile (ascii csv files, XML files, HTML tables) and is independent of the database driver. If in the future the database is to be translated  on more efficient SQL servers (which may not be MySQL) the requests and the answer will always remain the same from the end-user point of view.

Various projects, corresponding to different studies, are stored in the 3MdB. The parameters and the outputs of the models that are stored in the database do not depend on the project, but are kept the same for any entry, asserting that models from different projects have the same fields. It means that we do not only store the values of the outputs that were used for a given study that drove the creation of some project, but instead we keep all the values of all the parameters and outputs that have been defined as useful when designing the database. In particular, we store the intensities of 178 emission lines and continuum fluxes, even if we only consider a very few of them in each project. The same applies for the ionic fractions of all the ions.

The access to the 3MdB is free and we are welcoming any new project i. e. grid of photoionization models that would benefit being stored in the 3MdB. In the present stage of 3MdB development, only results obtained from {\sc CLOUDY} runs are acceptable, but in a second stage results from other photoionization codes like MAPPINGS \citep{1993Sutherland_apjs88, 2013Dopita_apjs208} or MOCASSIN \citep{2003Ercolano_mnra340} could be implemented.

In the first Section, we present the structure of the 3MdB. The second section is devoted to the description of the projects currently being held by the 3MdB. In the third section we present some applications of the 3MdB. These examples must not be considered as  complete and detailled scientific studies, but  rather  as show-cases on how the 3MdB can be used. 

\section{The structure of the 3MdB}
\label{sec:3MdB}

\subsection{Using MySQL}

The 3MdB database is stored in a MySQL relational database management system. This can be seen as huge spreadsheets {\it a la} Microsoft Excel. It actually contains 5 tables (like 5 sheets): the main table, the ionic fraction table, the ionic temperature table, and the emission line temperature table, respectively named ``tab", ``abion", ``teion", and ``temis" (see Sec.~\ref{sec:tables} for details), and one additional table (``lines") used to describe the emission lines. 

Each entry in any of the 4 tables (like each row in a spreadsheet) corresponds to a photoionization model, and each column to a variable (a field, like the parameters of the models and some results of the run). When results for a given model are needed from more than one table (e. g. emission line intensities and ionic fractions), a unique reference number is used to link the tables using the joining capacities of MySQL.  

The MySQL language allows the user to search for values of fields corresponding to entries fitting a given criterium or a combination of different criteria. The criteria can be any combination of logical test on the values of any field (e.g. looking for all the models of a given project with [\ion{O}{iii}]/H$\beta$ = 1.5 within 10\% and [\ion{N}{ii}]/H$\alpha$ = 1.2 within 15\%, models using a BlackBody as ionizing source and with log(U)\footnote{$U = \frac{Q_0}{4.\pi.r^2.N_e.c}$ with $Q_0$ the number of ionizing photons emitted by the central source, $r$ the distance between the source and the gas, and $N_e$ the electron density, $c$ being the speed of the light. The 3MdB stores log(U) at the first step, at the last step and its mean value over the volume of the nebula, weighted by ne.nH.}  between $-3.5$ and $-1.5$; from these models returns some selected line ratios and ionic fractions). 
The power of MySQL to treat the requests in a few seconds (to a few minutes for very complexe criteria applied on numerous entries) gives new possibilities of exploration and of data-mining. To speed up the requests, some fields have been set up as indices so that they can be accessed faster (see Tab.~\ref{tab:tab1}).

The whole system used to compute grids of models and to insert the results into the different tables of the database has been included as part of the pyCloudy package \citep{2013Morisset_, 2014Morisset_}.

\subsection{The fields of the different tables}
\label{sec:tables}
\begin{itemize}
\item {\bf tab} This table contains all the input parameters of the model and some of the outputs. 
The fields of the table ``tab'' are described in Tab.~\ref{tab:tab1} for the inputs of the run, and Tab.~\ref{tab:tab2} for some of the outputs. All the options needed to run the code are stored in the 3MdB, so that any {\sc CLOUDY} input file can be generated from the 3MdB data for the models to be rerun.
In addition, the table contains the abundances (in log(X/H)) of the following elements: 
Hydrogen,
 Helium,
 Lithium,
 Beryllium,
 Boron,
 Carbon,
 Nitrogen,
 Oxygen,
 Fluorine,
 Neon,
 Sodium,
 Magnesium,
 Aluminium,
 Silicon,
 Phosphorus,
 Sulphur,
 Chlorine,
 Argon,
 Potassium,
 Calcium,
 Scandium,
 Titanium,
 Vanadium,
 Chromium,
 Manganese,
 Iron,
 Cobalt,
 Nickel,
 Copper,
and Zinc.

The ionizing SED (Spectral Energy Distribution) can be defined as the sum of 2 spectra (Blackbody, classical atmosphere models provided by {\sc CLOUDY} or user-defined SED). If more than 2 spectra are needed, a call to an extra table containing as much SEDs as needed can be invoqued (this is not treated in this paper, as the projects currently held in 3MdB do not use this facility).

Line intensities are also part of the ``tab" table. All the lines are given in Tab.~\ref{tab:lines}. Notice that for every line, the field name  (e. g. ``O\_\_3\_\_5007A") refers to the volume integral of the line emissivity, while the field name ending with \_rad (e. g. ``O\_\_3\_\_5007A\_rad") refers to the line emissivity integrated over the radius. The intensities are given in $\rm erg\,s^{-1}$.
The table contains 508 fields. 
\item {\bf abion} This table contains the ionic fractions for the 492 ions computed by {\sc CLOUDY}. The corresponding field names are, for example for O$^{++}$: ``A\_OXYGEN\_vol\_2" and "A\_OXYGEN\_rad\_2", depending on whether the integral is performed over the radius or the volume. Here is an example in the case of volume integration: 
$$\frac{\int O^{++}/O.n_e.n_H.ff.dV}{\int n_e.n_H.ff.dV}$$
where $O^{++}/O$ is the ionic fraction of ion $O^{++}$, $n_e$ and $n_H$ are the electron and hydrogen densities respectively, and $ff$ is the filling factor. The table contains 994 fields. 
\item {\bf teion} This table contains the electron temperature weighted by the ionic fractions for the 492 ions computed by {\sc CLOUDY}. The corresponding field names are, for example for O$^{++}$: ``T\_OXYGEN\_vol\_2" and ``T\_OXYGEN\_rad\_2", depending on whether the integral is performed over the radius or the volume. Here is an example in the case of the volume integration:
$$\frac{\int T_e.O^{++}/O.n_e.n_H.ff.dV}{\int O^{++}/O.n_e.n_H.ff.dV}$$
The table contains the same 994 fields than ``abion''. 
\item {\bf temis} This table contains the electron temperature weighted by the line emissivity, for all the lines described in Tab.~ 3. For example, for the $[\ion{O}{iii}]\lambda$ 5007\AA, the field name is ``T\_O\_\_3\_\_5007A" and the result corresponds to:
$$\frac{\int T_e.\epsilon([\ion{O}{iii}]\lambda 5007).dV}{\int \epsilon([\ion{O}{iii}]\lambda 5007).dV}$$
where $\epsilon([\ion{O}{iii}]\lambda 5007)$ is the line emissivity as given by {\sc CLOUDY}. The table contains 180 fields (178 for the temperatures, the table also includes the ``N'' and ``ref'' fields to easily join it to other tables).
\end{itemize}

\begin{table*}
\caption{Fields of the table ``tab'' (1st part). Stars denote MySQL indexes.} 
\label{tab:tab1} 
\begin{tabular}{c|l}
\toprule
field name & description \\ \midrule  \midrule
N & unique model identifier* \\  \midrule
user & name of the user who owns the project*\\  \midrule
ref & name of the project* \\  \midrule
file & generic filename of the model\\  \midrule
dir & directory where the model files are stored \\  \midrule
C\_version& {\sc CLOUDY} version (e. g. ``Cloudy 13.03'')\\  \midrule
geom& geometry (sphere or Null)\\  \midrule
atm\_cmd& BlackBody, or table star ``xxx.mod''. 1st SED\\  \midrule
atm\_file& name of the atmosphere file, if not in atm\_cmd. 1st SED*\\  \midrule
atm1& 1st parameter for the atmosphere file. 1st SED\\  \midrule
atm2& 2nd parameter for the atmosphere file. 1st SED\\  \midrule
atm3& 3rd parameter for the atmosphere file. 1st SED\\  \midrule
lumi\_unit& Luminosity unit (string used by {\sc CLOUDY}, e.g. ``q(H)''). 1st SED \\  \midrule
lumi& Value of the luminosity. 1st SED\\  \midrule
atm\_cmd2& BlackBody, or table star ``xxx.mod''. 2nd SED\\  \midrule
atm\_file2& name of the atmosphere file, if not in atm\_cmd. 2nd SED\\  \midrule
atm12& 1st parameter for the atmosphere file. 2nd  SED\\  \midrule
atm22& 2nd parameter for the atmosphere file. 2nd SED\\  \midrule
atm32& 3rd parameter for the atmosphere file. 2nd SED\\  \midrule
lumi\_unit2& Luminosity unit (string used by {\sc CLOUDY}, e.g. ``q(H)''). 2nd SED \\  \midrule
lumi2& Value of the luminosity. 2nd SED\\  \midrule
dens& density (in case of constant density model) [log cm$^{-3}$]\\  \midrule
dlaw1, dlaw2, ..., dlaw9& parameters for the density law\\  \midrule
radius& inner radius of the nebula [log cm]\\  \midrule
ff& filling factor\\  \midrule
dust\_type1, .., dust\_type3& types of the dust\\  \midrule
dust\_value1, ..., dust\_value3& values of the dust abundance\\  \midrule
stop1, ..., stop6& stopping criteria\\  \midrule
cloudy1, ..., cloudy9& additional {\sc CLOUDY} commands\\  \midrule
com1, ..., com9& comments \\  \midrule
distance& distance to the object [kpc]\\  \midrule
N\_Mass\_cut, N\_Hb\_cut& values used for the matter-bounded models\\  \midrule
precursor, generation &  used in the genetic algorithms  \\ 
\bottomrule
\end{tabular}
\end{table*}

\begin{table*}
\caption{Fields of the table ``tab'' (2nd part). } 
\label{tab:tab2}
\begin{tabular}{c|l}
\toprule
field name & description \\ \midrule  \midrule
depthFrac, massFrac, HbFrac& depth, mass and Hbeta fraction relative to the radiation bounded model\\  \midrule
rout& Outer radius [log cm]\\  \midrule
thickness& Thickness [cm]\\  \midrule
N\_zones& Number of zones\\  \midrule
CloudyEnds& the copy of the output from {\sc CLOUDY} where is says how it ends\\  \midrule
FirstZone& the description of the first zone, from the {\sc CLOUDY} output\\  \midrule
LastZone& the description of the last zone, from the {\sc CLOUDY} output\\  \midrule
CalculStop& the copy of the output from {\sc CLOUDY} where is says why it stopped\\  \midrule
logQ& log(Q), where Q is the number of ionizing photons emitted per sec [log s$^{-1}$]\\  \midrule
logQ0& Q0 is the number of ionizing photons between 1 and 1.807 Ryd [log s$^{-1}$]\\  \midrule
logQ1& Q1 is the number of ionizing photons between 1.807 and 4 Ryd [log s$^{-1}$]\\  \midrule
logQ2& Q2 is the number of ionizing photons between 4 and 20.6 eV [log s$^{-1}$]\\  \midrule
logQ3& Q3 is the number of ionizing photons between 20.6 and 7676 eV [log s$^{-1}$]\\  \midrule
logPhi& log(Phi), where Phi is Q0 per surface unit [log s$^{-1}$.cm$^{-2}$]\\  \midrule
logPhi0, logPhi1, logPhi2, logPhi3& same as for Q0, Q1, Q2 Q3 but for Phi [log s$^{-1}$.cm$^{-2}$]\\  \midrule
LogU\_in& log(U) at the inner radius\\  \midrule
LogU\_out& log(U) at the outer radius\\  \midrule
LogU\_mean& log(U) mean on the volume weighted by {$n_e.n_H$}\\  \midrule
t2\_H1, t2\_O1, t2\_O2, t2\_O3& $t^2$ {\it a la} Peimbert\\  \midrule
ne\_H1, ne\_O1, ne\_O2, ne\_O3& electron densities, weighted by ionic abundances [cm$^{-3}$]\\  \midrule
H\_mass, H1\_mass& Hydrogen masses (H total, H$^+$) [Msol]\\  \midrule
nH\_in, nH\_out, nH\_mean& Hydrogen density at the inner and outer radius, and mean over the volume\\  \midrule
Hb\_SB& H$\beta$ surface brightness\\  \midrule
Hb\_EW, Ha\_EW& H$\beta$ and H$\alpha$ equivalent width\\  \midrule
Cloudy\_version& {\sc CLOUDY} version\\  \midrule
interpol& 1 if the model is interpolated between two complete {\sc CLOUDY} models\\  \midrule
datetime& date and time when the model is included in the 3MdB\\  \midrule
N1, N2, W1& used in case of N1 and N2 combined models with weight W1\\  
\bottomrule
\end{tabular}
\end{table*}

\begin{table*}
\caption{The 178 emission lines and continuum held in the 3MdB. The 3MdB columns give the field name in the 3MdB. The {\sc CLOUDY} columns refer to the command line used in the input of the photoionization code.} 
\label{tab:lines} 
\scalebox{0.7}{
\begin{tabular}{r|r|r||r|r|r||r|r|r}
\toprule
3MdB & {\sc CLOUDY} & Comments & 3MdB & {\sc CLOUDY} & Comments & 3MdB & {\sc CLOUDY} & Comments \\
\midrule
BAC\_\_\_3646A & Bac    3646 & BalmHead & COUT\_\_3646A & cout   3646 & OutwardBalmPeak & CREF\_\_3646A & cref   3646 & ReflectedBalmPeak \\ 
H\_\_1\_\_4861A & H  1   4861 & H I 4861 & TOTL\_\_4861A & TOTL   4861 & H I 4861 & H\_\_1\_\_6563A & H  1   6563 & H I 6563 \\ 
H\_\_1\_\_4340A & H  1   4340 & H I 4340 & H\_\_1\_\_4102A & H  1   4102 & H I 4102 & H\_\_1\_\_3970A & H  1   3970 & H I 3970 \\ 
H\_\_1\_\_3835A & H  1   3835 & H I 3835 & H\_\_1\_\_1216A & H  1   1216 & H I 1216 & H\_\_1\_4051M & H  1  4.051 & H I 4.051m \\ 
H\_\_1\_2625M & H  1  2.625 & H I 2.625m & H\_\_1\_7458M & H  1  7.458 & H I 7.458m & HE\_1\_\_5876A & He 1   5876 & He I 5876 \\ 
CA\_B\_\_5876A & Ca B   5876 & He I 5876 Bcase & HE\_1\_\_7281A & He 1   7281 & He I 7281 & HE\_1\_\_7065A & He 1   7065 & He I 7065 \\ 
HE\_1\_\_4471A & He 1   4471 & He I 4471 & CA\_B\_\_4471A & Ca B   4471 & He I 4471 Bcase & HE\_1\_\_6678A & He 1   6678 & He I 6678 \\ 
CA\_B\_\_6678A & Ca B   6678 & He I 6678 Bcase & TOTL\_1083M & TOTL  1.083 & He I 1.083 & HE\_2\_\_1640A & He 2   1640 & He I 1640 \\ 
HE\_2\_\_4686A & He 2   4686 & He II 4686 & C\_\_1\_\_8727A & C  1   8727 & [C I] 8727 & TOTL\_\_9850A & TOTL   9850 & [C I] 9850 \\ 
C\_IC\_\_9850A & C Ic   9850 & [C I] 9850 coll & TOTL\_\_2326A & TOTL   2326 & C II] 2326+ & C\_\_2\_\_1335A & C  2   1335 & C II 1335 \\ 
C\_\_2\_\_1761A & C  2   1761 & C II 1761 & TOTL\_\_6580A & TOTL   6580 & [C II] 6580 & C\_\_2\_\_4267A & C  2   4267 & C II 4267 \\ 
C\_\_2\_1576M & C  2  157.6 & [C II] 157.6m & C\_\_3\_9770A & C  3  977.0 & [C III] 977 & C\_\_3\_\_1907A & C  3   1907 & [C III] 1907 \\ 
C\_\_3\_\_1910A & C  3   1910 & [C III] 1910 & C\_\_3\_\_4649A & C  3   4649 & C III 4649 & C\_\_3\_\_2297A & C  3   2297 & C III 2297 \\ 
TOTL\_\_1549A & TOTL   1549 & C IV 1549 totl & C\_\_4\_\_1549A & C  4   1549 & C IV 1549 rec & C\_\_4\_\_4659A & C  4   4659 & C IV 4649 \\ 
N\_\_1\_\_5198A & N  1   5198 & [N I] 5198 & N\_\_1\_\_5200A & N  1   5200 & [N I] 5200 & N\_\_2\_\_5755A & N  2   5755 & [N II] 5755 \\ 
N\_2R\_\_5755A & N 2r   5755 & N II 5755 rec & N\_\_2\_\_6548A & N  2   6548 & [N II] 6548 & N\_\_2\_\_6584A & N  2   6584 & [N II] 6584 \\ 
N\_\_2\_\_2141A & N  2   2141 & N II 2141 & N\_\_2\_\_4239A & N  2   4239 & N II 4239 & N\_\_2\_\_4041A & N  2   4041 & N II 4041 \\ 
TOTL\_\_5679A & TOTL   5679 & N II 5679 totl & N\_\_2\_1217M & N  2  121.7 & [N II] 121.7m & N\_\_2\_2054M & N  2  205.4 & [N II] 205.4m \\ 
N\_\_3\_5721M & N  3  57.21 & [N III] 57.21m & N\_\_3\_\_4641A & N  3   4641 & N III 4641 & TOTL\_\_1750A & TOTL   1750 & N III] 1750+ \\ 
N\_\_3\_\_4379A & N  3   4379 & N III 4379 & N\_\_4\_\_1485A & N  4   1485 & N IV] 1485 & N\_\_4\_\_1719A & N  4   1719 & N IV 1719 \\ 
TOTL\_\_1240A & TOTL   1240 & [N V] 1240 totl & N\_\_5\_\_1239A & N  5   1239 & [N V] 1239 & O\_\_1\_\_7773A & O  1   7773 & O I 7773 \\ 
O\_\_1\_\_6300A & O  1   6300 & [O I] 6300 & O\_\_1\_\_5577A & O  1   5577 & [O I] 5577 & O\_\_1\_6317M & O  1  63.17 & [O I] 63.17m \\ 
O\_\_1\_1455M & O  1  145.5 & [O I] 145.5m & O\_II\_\_3726A & O II   3726 & [O II] 3726 & O\_II\_\_3729A & O II   3729 & [O II] 3729 \\ 
O\_II\_\_7323A & O II   7323 & [O II] 7323 & O\_II\_\_7332A & O II   7332 & [O II] 7332 & O\_2R\_\_3726A & O 2r   3726 & O II 3726 rec \\ 
O\_2R\_\_3729A & O 2r   3729 & O II 3729 rec & O\_2R\_\_7323A & O 2r   7323 & O II 7323 rec & O\_2R\_\_7332A & O 2r   7332 & O II 7332 rec \\ 
TOTL\_\_3727A & TOTL   3727 & [O II] 3727+ & TOTL\_\_7325A & TOTL   7325 & [O II] 7325+ & O\_II\_\_2471A & O II   2471 & [O II] 2471+ \\ 
O\_\_2\_\_4152A & O  2   4152 & O II 4152 & TOTL\_\_4341A & TOTL   4341 & O II 4341 & O\_\_2\_\_4651A & O  2   4651 & O II 4651 \\ 
O\_2R\_\_4651A & O 2r   4651 & O II 4651+ & TOTL\_\_4363A & TOTL   4363 & [O III] 4363 & REC\_\_\_4363A & Rec    4363 & O III 4363 rec \\ 
O\_\_3\_\_4959A & O  3   4959 & [O III] 4959 & O\_\_3\_\_5007A & O  3   5007 & [O III] 5007 & O\_\_3\_5180M & O  3  51.80 & [O III] 51.8m \\ 
O\_\_3\_8833M & O  3  88.33 & [O III] 88.33m & TOTL\_\_1665A & TOTL   1665 & [O III] 1665+ & TOTL\_\_1402A & TOTL   1402 & O IV] 1402+ \\ 
O\_\_4\_\_1342A & O  4   1342 & O IV 1342 & O\_\_4\_2588M & O  4  25.88 & [O IV] 25.88m & TOTL\_\_1218A & TOTL   1218 & O V] 1218+ \\ 
O\_\_5\_\_1216A & O  5   1216 & [O V] 1216 & NE\_2\_1281M & Ne 2  12.81 & [Ne II] 12.81m & NE\_3\_\_3869A & Ne 3   3869 & [Ne III] 3869 \\ 
NE\_3\_\_3968A & Ne 3   3968 & [Ne III] 3968 & NE\_3\_1555M & Ne 3  15.55 & [Ne III] 15.55m & NE\_3\_3601M & Ne 3  36.01 & [Ne III] 36.01m \\ 
NE\_3\_\_1815A & Ne 3   1815 & [Ne III] 1815 & NE\_4\_\_1602A & Ne 4   1602 & [Ne IV] 1602 & NE\_4\_\_2424A & Ne 4   2424 & [Ne IV] 2424 \\ 
NE\_4\_\_4720A & Ne 4   4720 & [Ne IV] 4720+ & NE\_5\_\_3426A & Ne 5   3426 & [Ne V] 3426 & NE\_5\_\_3346A & Ne 5   3346 & [Ne V] 3346 \\ 
NE\_5\_\_2976A & Ne 5   2976 & [Ne V] 2976 & NE\_5\_2431M & Ne 5  24.31 & [Ne V] 24.31m & NE\_5\_1432M & Ne 5  14.32 & [Ne V] 14.32m \\ 
TOTL\_\_2798A & TOTL   2798 & [Mg II] 2798+ & SI\_2\_3481M & Si 2  34.81 & [Si II] 34.81m & SI\_2\_\_2334A & Si 2   2334 & [Si II] 2334 \\ 
SI\_3\_\_1892A & Si 3   1892 & [Si III] 1892 & SI\_4\_\_1394A & Si 4   1394 & [Si IV] 1394 & S\_II\_\_4070A & S II   4070 & [S II] 4070 \\ 
S\_II\_\_4078A & S II   4078 & [S II] 4078 & S\_II\_\_6731A & S II   6731 & [S II] 6731 & S\_II\_\_6716A & S II   6716 & [S II] 6716 \\ 
S\_II\_1029M & S II  1.029 & [S II 1.029m & S\_II\_1034M & S II  1.034 & [S II] 1.034m & S\_II\_1032M & S II  1.032 & [S II] 1.032m \\ 
S\_II\_1037M & S II  1.037 & [S II] 1.037m & S\_\_3\_\_6312A & S  3   6312 & [S III] 6312 & S\_\_3\_\_9532A & S  3   9532 & [S III] 9532 \\ 
S\_\_3\_\_9069A & S  3   9069 & [S III] 9069 & S\_\_3\_1867M & S  3  18.67 & [S III] 18.67m & S\_\_3\_3347M & S  3  33.47 & [S III] 33.47m \\ 
S\_\_4\_1051M & S  4  10.51 & [S IV] 10.51m & S\_\_4\_\_1398A & S  4   1398 & [S IV] 1398 & CL\_2\_\_8579A & Cl 2   8579 & [Cl II] 8579 \\ 
CL\_2\_\_9124A & Cl 2   9124 & [Cl II] 9124 & CL\_2\_\_6162A & Cl 2   6162 & [Cl II] 6162 & CL\_2\_1440M & Cl 2  14.40 & [Cl II] 14.40m \\ 
CL\_3\_\_8552A & Cl 3   8552 & [Cl III] 8552 & TOTL\_\_8494A & TOTL   8494 & [Cl III] 8494+ & CL\_3\_\_5538A & Cl 3   5538 & [Cl III] 5538 \\ 
CL\_3\_\_5518A & Cl 3   5518 & [Cl III] 5518 & CL\_4\_\_7532A & Cl 4   7532 & [Cl IV] 7532 & CL\_4\_2040M & Cl 4  20.40 & [Cl IV] 20.40m \\ 
CL\_4\_1170M & Cl 4  11.70 & [Cl IV] 22.70m & AR\_2\_6980M & Ar 2  6.980 & [Ar II] 6.98m & AR\_3\_\_7135A & Ar 3   7135 & [Ar III] 7135 \\ 
AR\_3\_\_7751A & Ar 3   7751 & [Ar III] 7751 & AR\_3\_\_5192A & Ar 3   5192 & [Ar III] 5192 & AR\_3\_9000M & Ar 3  9.000 & [Ar III] 9.00m \\ 
AR\_3\_2183M & Ar 3  21.83 & [Ar III] 21.83m & AR\_4\_\_7171A & Ar 4   7171 & [Ar IV] 7171 & AR\_4\_\_4711A & Ar 4   4711 & [Ar IV] 4711 \\ 
AR\_4\_\_4740A & Ar 4   4740 & [Ar IV] 4740 & AR\_5\_\_7005A & Ar 5   7005 & [Ar V] 7005 & AR\_5\_1310M & Ar 5  13.10 & [Ar V] 13.1m \\ 
AR\_5\_8000M & Ar 5  8.000 & [Ar V] 8.00m & FE\_2\_\_8617A & Fe 2   8617 & [Fe II] 8617 & FE\_3\_\_4608A & Fe 3   4608 & [Fe III] 4608 \\ 
FE\_3\_\_4668A & Fe 3   4668 & [Fe III] 4668 & FE\_3\_\_4659A & Fe 3   4659 & [Fe III] 4659 & FE\_3\_\_4702A & Fe 3   4702 & [Fe III] 4702 \\ 
FE\_3\_\_4734A & Fe 3   4734 & [Fe III] 4734 & FE\_3\_\_4881A & Fe 3   4881 & [Fe III] 4881 & FE\_3\_\_5271A & Fe 3   5271 & [Fe III] 5271 \\ 
FE\_3\_\_4755A & Fe 3   4755 & [Fe III] 4755 & FE\_4\_\_2836A & Fe 4   2836 & [Fe IV] 2836 & FE\_6\_\_5177A & Fe 6   5177 & [Fe VI] 5177 \\ 
FE\_7\_\_4894A & Fe 7   4894 & [Fe VII] 4894 & FE\_7\_\_5721A & Fe 7   5721 & [Fe VII] 5721 & FE\_7\_\_4989A & Fe 7   4989 & [Fe VII] 4989 \\ 
FE\_7\_\_6087A & Fe 7   6087 & [Fe VII] 6087 & FE\_7\_\_5277A & Fe 7   5277 & [Fe VII] 5277 & F12\_\_1200M & F12  12.00 & IRAS 12m \\ 
F25\_\_2500M & F25  25.00 & IRAS 25m & F60\_\_6000M & F60  60.00 & IRAS 60m & F100\_1000M & F100  100.00 & IRAS 100m \\ 
MIPS\_2400M & MIPS  24.00 & MIPS 24m & MIPS\_7000M & MIPS  70.00 & MIPS 70m & MIPS\_1600M & MIPS  160.0 & MIPS 160m \\ 
IRAC\_3600M & IRAC  3.600 & IRAC 3.6m & IRAC\_4500M & IRAC  4.500 & IRAC 4.5m & IRAC\_5800M & IRAC  5.800 & IRAC 5.8m \\ 
IRAC\_8000M & IRAC  8.000 & IRAC 8.0m &  & & & & & \\ 
\bottomrule
\end{tabular}}
\end{table*}

\section{The projects currently held by the 3MdB}

There are currently 4 projects held in the 3MdB, but more will be added in the future. The 3MdB webpage\footnote{\url{https://sites.google.com/site/mexicanmillionmodels/}} is used to describe the projects currently in 3MdB as well as the forthcomming new ones. In the following sections, a description of the grids corresponding to each project is given.

\subsection{The DIG}
\label{sec:DIGs}

The Diffuse Ionized Gas (DIG) was detected through its optical line emission outside the classical \ion{H}{ii} regions \citep{1971Reynolds_} and turns out to be a major component of the interstellar medium in galaxies \citep{1991Reynolds_144}. An electron density $\rm \sim 0.1 \, cm^{-3}$ and temperature $\rm \sim 10^4 \, K$ are the physical conditions that caracterize the DIG. The main ionizing and heating sources of the DIG, were unknown for several decades. \citet{2011Flores-Fajardo_mnra415} have shown that photoionization models taking into account photons emittted by OB stars that have been leaked out of \ion{H}{ii} regions located in the galactic thin disk combined with photons coming from the expected population of HOt Low Mass Evolved Stars (HOLMES) are able to reproduce the emission-line features observed in edge-on galaxies.

We developed a grid of plane-parallel, ionization-bounded models included in the 3MdB under the reference ``DIG\_HR'' in which we combine two ionizations sources: one coming from the unabsorbed OB stars and one coming from the HOLMES. Unlike in the original grid, the one included in the 3MdB is a real high resolution grid in the sense that no interpolation between models is required, meaning that each point of the grid is an a independent and complete {\sc CLOUDY} model.
All the models are radiation bounded, as described in \citet{2011Flores-Fajardo_mnra415}.

In ``DIG\_HR'', the SED representing the OB stars is fixed and was developed with the evolutionary stellar population synthesis code {\sc STARBURST99} v 7.0.1 \citep{1999Leitherer_apjs123}, while the SED representing the HOLMES is fixed too and was developed with the evolutionary spectral synthesis code {\sc PEGASE.2} \citep{1997Fioc_aap326}, both with the detailed characteristics determined for the galaxy NGC891 described in \citet{2011Flores-Fajardo_mnra415}. In addition, each model is defined by the value of the OB surface flux ${\rm \Phi_{OB}}$, the ionization parameter $U$ and the chemical composition of the gas. The ionization parameter is defined as $U = {\rm \Phi_{total}} / (n_e c)$ where  ${\rm \Phi_{total} = \Phi_{OB} \, + \, \Phi_{HOLMES}}$, $n_e$ is the electron density and $c$ is the speed of light. The value of ${\rm \Phi_{HOLMES}} = 8.4 \times 10^4 \; {\rm photons \; s^{-1} \; cm^{-2}}$ is the same for all the models in this grid and once $\rm \Phi_{OB}$ is set, the value of $U$ defining the models are varied by changing the electron density $n_e$. Finally, the abundances of the heavy elements (except N) relative to O are fixed to their solar values as implemented in {\sc CLOUDY}. Mg, Si and Fe are depleted by 1 dex. While the original interpolated grid was dust free, the one included in 3MdB consider the mixture defined as ``ism'' by {\sc CLOUDY}. An additional parameter is the dust content in the models. While the original interpolated grid was dust free, the one included in 3MdB consider the mixture defined as ~ism~ by {\sc CLOUDY}.

The combination of all the parameters (see Table \ref{tab:DIG_params} for the ranges covered by the varying parameters) produces 41327 models computed and included in the 3MdB.

\begin{table*}
\caption{Varying parameters for the ref=``DIG\_HR'' models} 
\label{tab:DIG_params} 
\begin{tabular}{c|c|c|c|c|c}
\toprule
3MdB field name & description & lower value & higher value & steps & step number\\ \midrule   \midrule  
com1 & log(U) & $-4$ & $-3$ &  0.1 & 11\\ \midrule  
com2 & ${\rm \Phi_{OB}}$ & 3.5 & 7.5 & 0.25 & 17 \\ \midrule  
com3 & log N/O & $-1.4$ & $-0.2$ & 0.1 & 13  \\ \midrule  
com4 & delta O/H & $-1$ & 0.6 & 0.1 & 17  \\ 
\bottomrule
\end{tabular}
\end{table*}

\subsection{Planetary Nebulae}
\label{sec:PNe}
The grid of photoionization models included in the 3MdB under the reference ``PNe\_2014'' is described in detail in \citet{2014Delgado-Inglada_mnra440}. This grid was originally created to compute ionization correction factors (ICFs) for PNe but, since it covers a wide range of physical parameters, it can be used for 
many other purposes. 

Various families of models were computed by changing some of the initial assumptions. The spectral energy distribution of the ionizing stars (column 'com 1' in ``tab'') is either a blackbody ('BB') or a Rauch NLTE model atmosphere \citep[TR, see ][]{1997Rauch_aap320}. 
The density distribution of the gas (column 'com 2' in ``tab'') is constant (C) or follows a Gaussian law (G). Some of the models are radiation bounded 
(R) and others have been created by trimming the radiation bounded model at a certain percentage of the total gas mass. For example M20 in 'com3' means that the model is matter bounded and corresponds to 20\% of the of the mass of the radiation bounded model, M40 corresponds to 40\%, M60 to 60\%, and M80 to 80 \% respectively. The input metallicity of the models is indicated in column 'com4'. Some models have the default PN abundances in {\sc CLOUDY} (S). The low metallicity models (L) have half solar abundances, the high metallicity models (H) have twice solar abundances, and the very high metallicity models (VH) have four times solar abundances. Finally, the presence of dust in the nebulae is also a variable in the models. If the 'com5' key is 'D', dust is included in the model, otherwise there are no dust grains. The grains considered are a typical mixture of silicate and carbonates, as defined by the ``ism" option in {\sc Cloudy}.

Within each of the above families of models, we varied the input parameters in the ranges of values defined in Tab.~\ref{tab:PNe_params}. 

Taking all the above possibilities into account, we computed 108864 photoionization models (108590 are effectively in the 3MdB, some of the parameters combination leading to not converged models), leading to 542950 models in the 3MdB once each run has been cut to obtain matter-bounded models. In \citet{2014Delgado-Inglada_mnra440} we applied several filters to the grid to exclude those models without physical sense. For example, we exclude the models with a combination of $T_{\rm eff}$ and $L_*$ falling outside the typical evolutionary tracks (for example those from \citealt{1983Schoenberner_apj272, 1995Bloecker_aap299}), and those with hydrogen masses above 1 $M_\odot$ because higher masses are not observed \citep[e.g.][]{1987Barlow_mnra227, 1987Gathier_aaps71, 1991Stasinska_aap247}. 
The other criteria are described in \citet{2014Delgado-Inglada_mnra440} and hereafter. The com6 = 1 condition is filled when {\bf all} the following conditions are filled:
\begin{itemize}
\item L$_* < 4.2$ and (L$_* > 3.4$ or T$_ * > 100,000$~K) and L$_* > (1.5 \times 10^{-5} T_* - 0.25)$, where L$_* = log(L/L_\odot)$,
\item Mass$_H < 1 M_{\odot}$,
\item $N_H  R_{out}^3 > 2\times 10^{53}$ and $N_H  R_{out}^3 < 3\times 10^{56}$,
\item $-15 < log H_\beta Surface Brightness < -11$,
\item MassFrac = M(H) / M(H)$_{rad-bounded}$ $>$ 20\%
\end{itemize}

The column 'com6' in the table ``tab'' is 'NULL' if the model is rejected from our filters and '1' if the model passes our selection criteria. The number of models with com6 = 1 (the ones corresponding to realistic nebulae) is 84237.

\begin{table*}
\caption{Varying parameters for the ref=``PNe\_2014'' models} 
\label{tab:PNe_params} 
\begin{tabular}{c|c|c|c|c}
\toprule
3MdB field name & description & lower value & higher value & step number\\ \midrule   \midrule  
com1 & SED form & 'BB': BlackBody & 'TR': T. Rauch & 2 \\ \midrule 
com2 & density law & 'C': constant & 'G': gaussian & 2 \\ \midrule 
com3 & mass cut (mat- and rad-bounded) &  \multicolumn{2}{c|}{M20, M40, M60, M80, and R} & 5 \\ \midrule 
com4 & log(O/H) & \multicolumn{2}{c|}{$-3.66$:'L', $-3.36$:'S', $-3.06$:'H', $-2.76$:'VH'} &  4 \\ \midrule 
com5 & dust & 'N': no dust & 'D': dust &  2 \\ \midrule 
atm1 & T$_*$ for com1='BB'& 25,000~K & 300,000~K & 12  \\ \midrule 
atm1 &  T$_*$ for com1='TR'& 50,000~K & 180,000~K &  6  \\ \midrule 
dens & H density & 30 & 3.10$^5$ & 9 \\ \midrule 
lumi & stellar luminosity & \multicolumn{2}{c|}{2e2, 1e3, 3e3, 5.6e3, 1e4, 1.78e4} &  6 \\ \midrule 
radius & inner radius & \multicolumn{2}{c|}{3e15, 1e16, 3e16, 1e17, 3e17, 1e18, 3e18} & 7  \\  
\bottomrule
\end{tabular}
\end{table*}

\subsection{\ion{H}{ii} Regions}
\label{sec:HII}
The grid of photoionization models included in the 3MdB under the reference ``HII\_CHIm'' is described in detail in \citet{2014Perez-Montero_mnra441}. It is a small grid of \ion{H}{ii} regions models, initially done to determine chemical abundances from Te-method. The ionizing SED is obtained from a PopStar model \citep{2009Molla_mnra398} of a single instantaneous burst with an age of 1~Myr. Two options for the Carbon abundance have been explored: C/H is assumed to follow O/H (with a constant values of log(C/O) = $-0.26$), or N/H (with a constant value of log(C/N) = 0.6). The values taken by the varying parameters are given in Tab.~\ref{tab:HII_params}. The total number of models is then 11 x 21 x 17 x 2 = 7854.

\begin{table*}
\caption{Varying parameters for the ref=``HII\_CHIm'' models} 
\label{tab:HII_params} 
\begin{tabular}{c|c|c|c|c|c}
\toprule
3MdB field name & description & lower value & higer value & steps & step number\\ \midrule   \midrule  
lumi & log(U) & $-4$ & $-1.5$ & 0.25 & 11 \\ \midrule  
12 + oxygen & 12 + log O/H & 7.1 & 9.1 & 0.1 & 21  \\ \midrule  
nitrogen $-$ oxygen & log (N/O) & 0.0 & 2.0 & 0.125 & 17  \\ \midrule  
com1 & C following N or O & 'O' & 'N' & -- & 2  \\ \midrule
carbon $-$ nitrogen & log(C/N) when com1 = 'N' & 0.6 & 0.6 & --& 1  \\ \midrule
carbon $-$ oxygen & log(C/O) when com1 = 'O' & $-0.26$ & $-0.26$ & --& 1  \\ 

\bottomrule
\end{tabular}
\end{table*}

\subsection{CALIFA \ion{H}{ii} regions}
\label{sec:CALIFA}

The models under the ``CALIFA'' reference in the 3MdB correspond to a grid of models performed using the Starlight spectral base of simple stellar populations (SSPs) comprising four metallicities (Z = 0.2, 0.4, 1, and 1.5 solar metallicity), and 39 ages between t = 10$^6$ and $1.4\times 10^{10}$ yr. This base corresponds to the model-set ``GM'' described by \citet{2014Cid-Fernandes_aap561}. It is the base used in the analysis of the CALIFA observations \citep{2013Cid-Fernandes_aap557}.
We compute the ionizing SEDs corresponding to these metallicities and ages by interpolating in the PopStar \citep{2009Molla_mnra398} public grid of models.

The values of the varying parameters for this project are sumarized in Tab.~\ref{tab:CALIFA_params}. Two morphologies have been used (thick and thin models). The same metallicity is used for the ionizing source and for the ionized gas. Once the photoionization models are computed, we store in the 3MdB the results corresponding to 20\%, 40\%, 60\%, 80\%, and 100\% of the mass of the radiation-bounded models. Dust is included following the \citet{2014Remy-Ruyer_aap563} broken law and adding a factor of 2/3 to the dust to gas ratio, following \citet{2011Draine_apj732}.

This leads to a grid of 39 x 4 x 11 x 5 x 2 x 5 = 85800 entries in the 3MdB. More details on the model parameters are available on the 3MdB webpage.

\begin{table*}
\caption{Varying parameters for the ref=``CALIFA'' models} 
\label{tab:CALIFA_params} 
\begin{tabular}{c|c|c|c|c|c}
\toprule
3MdB field name & description & lower value & higher value & steps & step number\\ \midrule   \midrule  
com1 & log(U) & $-4$ & $-1.5$ & 0.25 & 11 \\ \midrule  
com2 & form factor & 0.03 & 3.00 & see text & 2 \\ \midrule 
com3 & age & 10$^6$ & 1.4  10$^{10}$ & see \citet{2013Cid-Fernandes_aap557} & 39 \\ \midrule 
com4 & metallicity [solar] & 0.2 & 1.5 & see text & 4 \\ \midrule 
com5 & log N/O & $-0.5$ & 0.5 & 0.25 & 5 \\ \midrule
HbFrac & cut in H$\beta$/H$\beta$total & $\sim$20\% & $\sim$100\% & $\sim$20\% & 5 \\
\bottomrule
\end{tabular}
\end{table*}

\section{Examples of use}

\subsection{On the validity of electron temperature diagnostics}

\begin{figure*}\centering
\includegraphics[width=17cm]{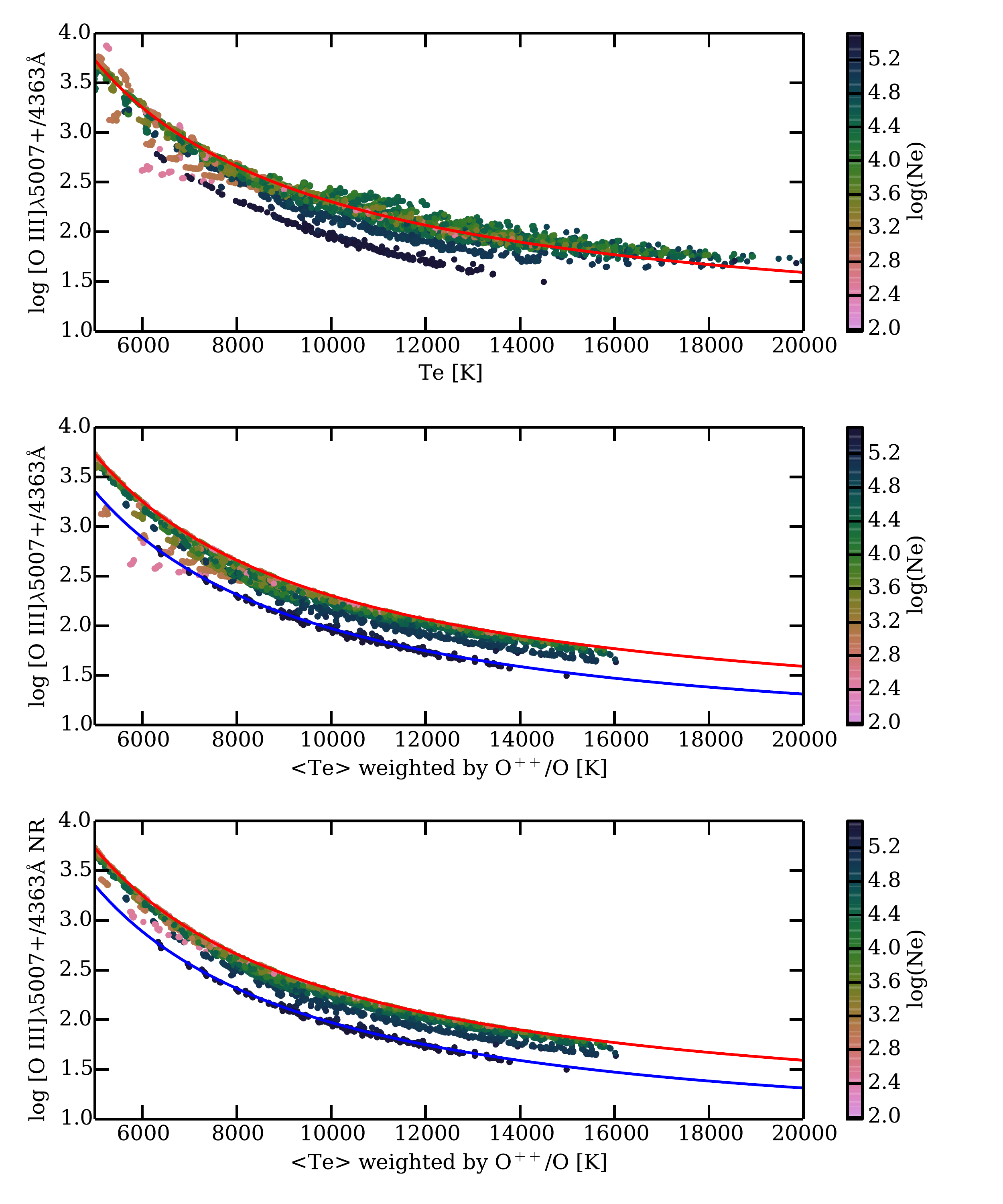}

\caption{Temperature determined by  $[\ion{O}{iii}]\lambda$5007+4959/4363\AA\  line ratio (using PyNeb), vs. electron temperature. The color codes the electron density. Upper panel: the electron temperature is the mean over the whole nebula. Middle panel: the electron temperature is weighted by O$^{++}$/O, thus tracing the O$^{++}$ region. Lower panel: the $[\ion{O}{iii}]\lambda$5007+4959/4363\AA\  line ratio is not taking into account the contribution of the recombination to $[\ion{O}{iii}]\lambda$4363\AA.  The red (upper) and the blue (lower) lines are theoretical values, corresponding to the low density limit and to a density of $3.10^{5}$ cm$^{-3}$ respectively.
\label{fig:Tdiag} }
\end{figure*}

The 3MdB can be used for teaching purposes, to demonstrate the validity of some assumptions or the relative orders of magnitude of different effects on the intensities of emission lines. To illustrate this, we plot in  Fig.~\ref{fig:Tdiag} the $[\ion{O}{iii}]\lambda$5007+4959/4363\AA\ line ratio versus the electron temperature of the gas, for a subset\footnote{We select the models which are radiation-bounded, ionized by a BlackBody SED, and without dust. We also apply the selection criteria com6 = 1 (see Sec.~\ref{sec:PNe})} of models under the ``PNe\_2014'' reference (see Sec.~\ref{sec:PNe}).
The color code indicates the electron density of the models. One can see in the upper panel that the models follow the red line, which draws the theoretical ratio as determined by PyNeb \citep{2014Luridiana_ArXi} at low density limit. Nevertheless, the scatter around this line, above and below it, seems larger than what one would expect for this diagnostic ratio. To understand this scatter, we then plot in the middle panel the same line ratio but versus the electron temperature weighted by the ionic fraction of O$^{++}$ (this is one of the fields of the 3MdB). This temperature actually corresponds to the region emitting the lines used for the diagnostics, thus leading to smaller scatter in $[\ion{O}{iii}]\lambda$5007+4959/4363\AA: the scatter is now only due to models with a lower value of the diagnostic than the theoretical one. The remaining scatter can be seen as due to two different causes: 
\begin{itemize} 
\item The effect of the collisional de-excitation of the level 4 where the $[\ion{O}{iii}]\lambda$5007+4959\AA\ are coming from (the critical densities of level 4 and 5 at 10,000~K are 7.10$^5$ cm$^{-3}$ and 2.10$^7$ cm$^{-3}$ resp.). This is clearly illustrated by the color gradient in the scatter, corresponding to the electron density and can be reproduced by increasing the density in the computation of the theoretical value of the line ratio. We can see the difference between the red and the blue lines (in the middle and lower panels), corresponding to the low density limit, and to a density of $3.10^{5}$ cm$^{-3}$ respectively. 
\item At low temperatures and low densities, another discrepancy occurs between the theoretical value and the one computed by the detailed photoionization model. This is due to the contribution of the recombination to the $[\ion{O}{iii}]\lambda$4363\AA\ intensity. This effect is increasing when the temperature decreases (following the increase of the emissivity of the recombination line) and is more important when the recombining ion (O$^{3+}$) is dominant, i. e. at low density (leading to higher ionization parameter) and high effective stellar temperature. By the way, this effect is what \citet{2014Proxauf_aap561} wrongly attribute to the fluorescence (absent in {\sc CLOUDY}) to interpret their results obtained with irrealistic models with huge ionization parameter \citep[see][]{2014Luridiana_ArXi}. This can be verified by plotting the line ratio subtracting from $[\ion{O}{iii}]\lambda$4363\AA\ the contribution of the recombination (which is available from {\sc CLOUDY} and recorded in the 3MdB). This is what is plotted in the lower panel, where the low density outsiders at low temperature disappear.
\end{itemize} 

\subsection{The effect of the ionization parameter on T$_e$([\ion{O}{iii}])}

\begin{figure}\centering
\includegraphics[width=8.2cm]{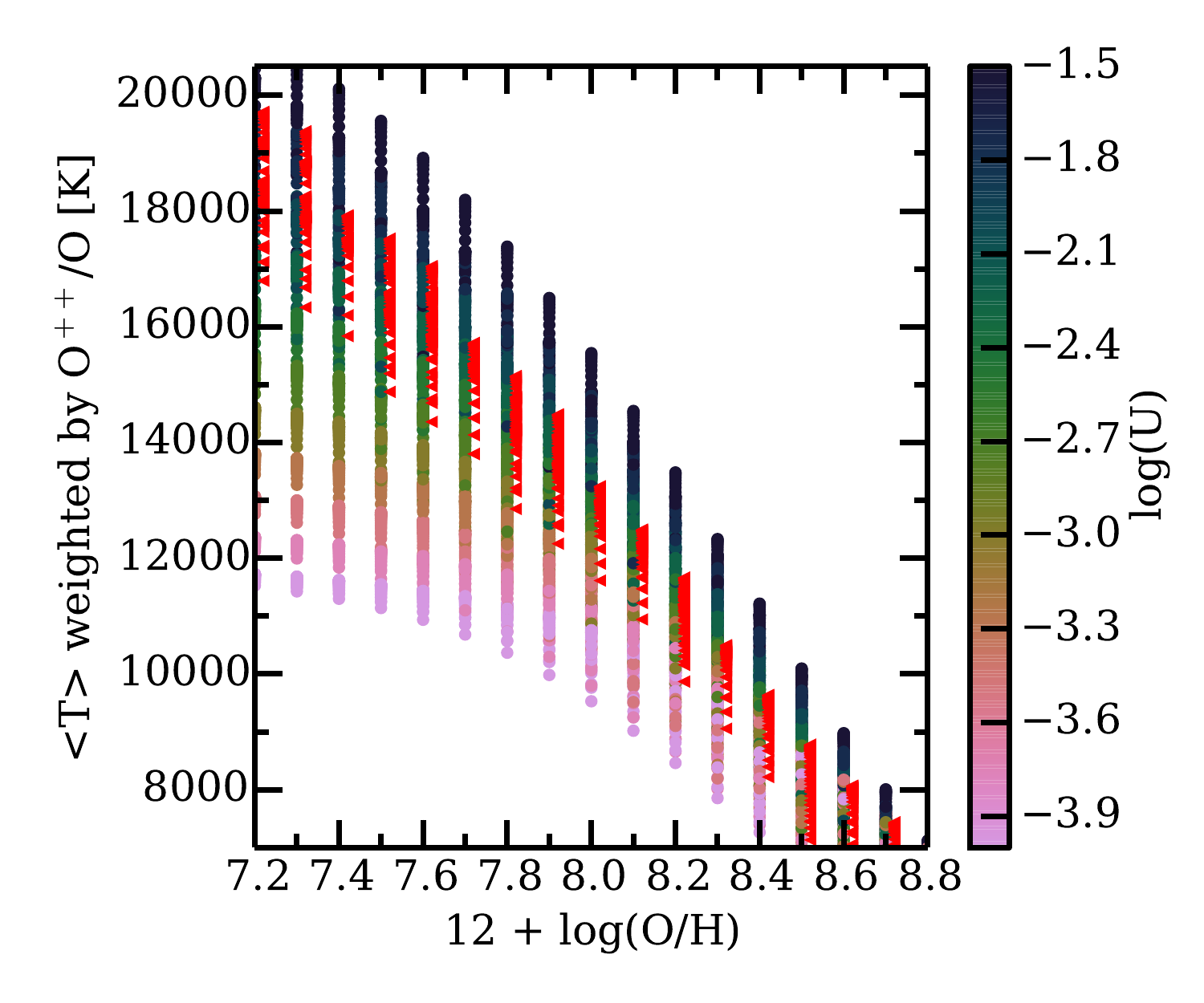}
\caption{Electron temperature of the O$^{++}$ region vs. Oxygen abundance, colored by logU. The data are taken from the 3MdB ``HII\_CHIm'' models. Red triangles (slightly x-shifted to be more visible) correspond to the models that fit the relation log(U) = $3.93 - 0.8 \times $(12 + log O/H), within a tolerance of 0.2 dex. \label{fig:TO_abO} }
\end{figure}

In a recent paper, \citet{2014Nicholls_apj790} compare the distribution of the [\ion{O}{iii}] electron temperature versus the oxygen gas-phase abundance observed in their sample of small isolated gas rich irregular dwarf galaxies (SIGRID). They find that classical simple photoionization models cannot well reproduce the observed behavior, in particular the high values of T([\ion{O}{iii}]) obtained at low metallicity. The authors explore the different scenarios that can actually fit these extreme observations, and concluded that a combination of pressure and optical depths could readily explain them.
We use here all the models from the 3MdB under the ``HII\_CHIm'' reference (see \ref{sec:HII} and Tab.~\ref{tab:HII_params}) and plot in Fig.~\ref{fig:TO_abO} the same quantities with the same axes ranges as in their Fig.~12. 
We find that we can reproduce all the observed points, once we take the ionization parameter log(U) as a free parameter. Using the relation log(U) $\propto Z^{-0.8}$  from \citet{2006Dopita_apj647}, normalized with log(U) = $-3$ at solar metallicity (i. e. 12 + log O/H = 8.69), we define a criterion to select the models so that log(U) = $3.93 - 0.8 \times $(12 + log O/H), within a tolerance of 0.2 dex. This relation is similar to the one used by \citet{2014Perez-Montero_mnra441}. Those models are plotted in red in the same figure, and fit quite well the observed points from the SIGRID sample, even the extreme ones at low metallicity.

\subsection{When HOLMES ionize \ion{H}{ii} regions}
\label{sec:Holmes}

\begin{figure}\centering
\includegraphics[width=8.2cm]{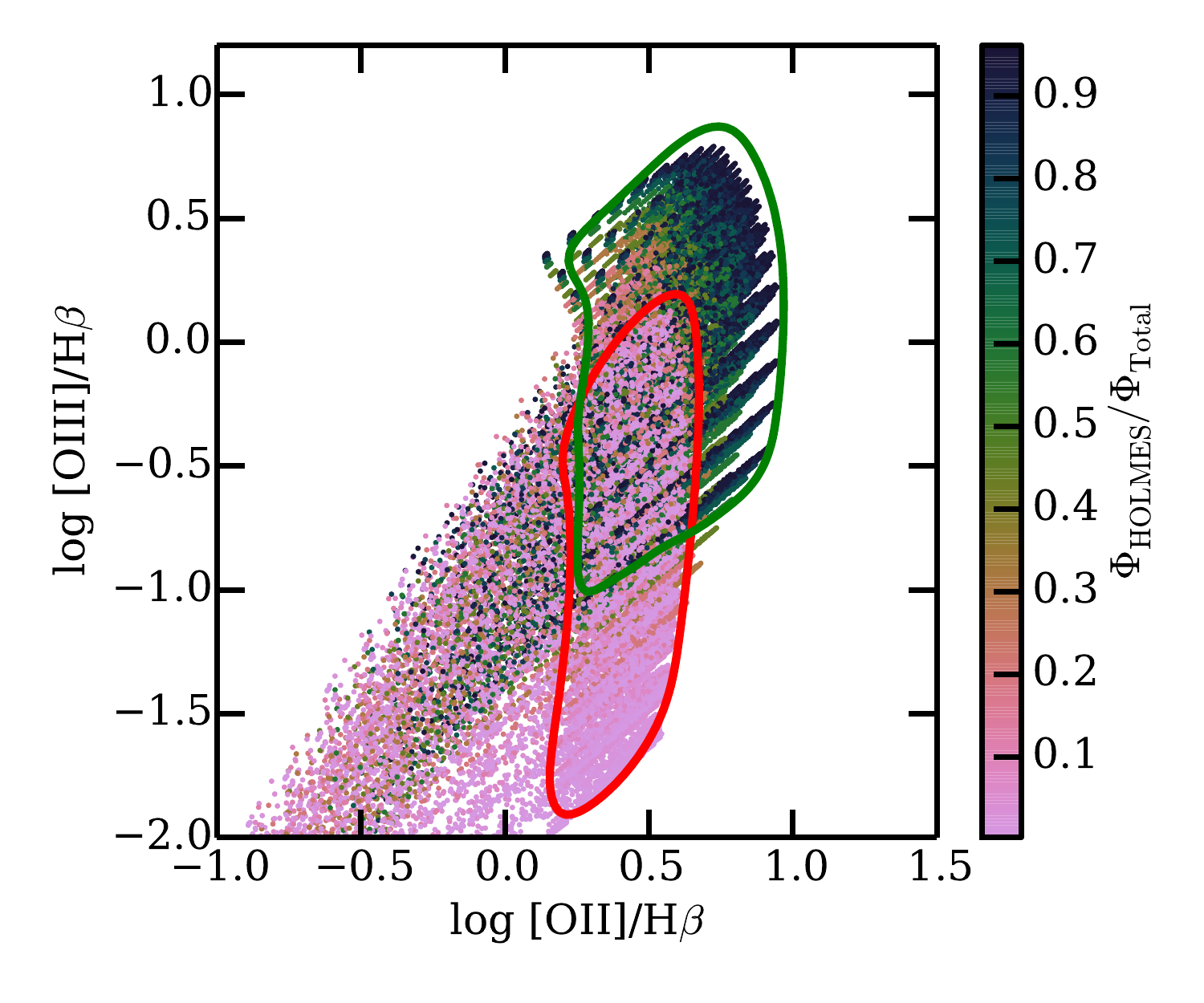}
\caption{[\ion{O}{iii}]/H$\beta$ vs. [\ion{O}{ii}]/H$\beta$ for models referenced as ``DIG\_HR''. The red contour (the lower one) contains half of the points for which the contribution of the OB stars is greater than 95\%. The green contour (the upper one) contains half of the points for which the contribution of the HOLMES is greater than 90\%.
 \label{fig:DIG1} }
\end{figure}

\begin{figure*}\centering
\includegraphics[width=16cm]{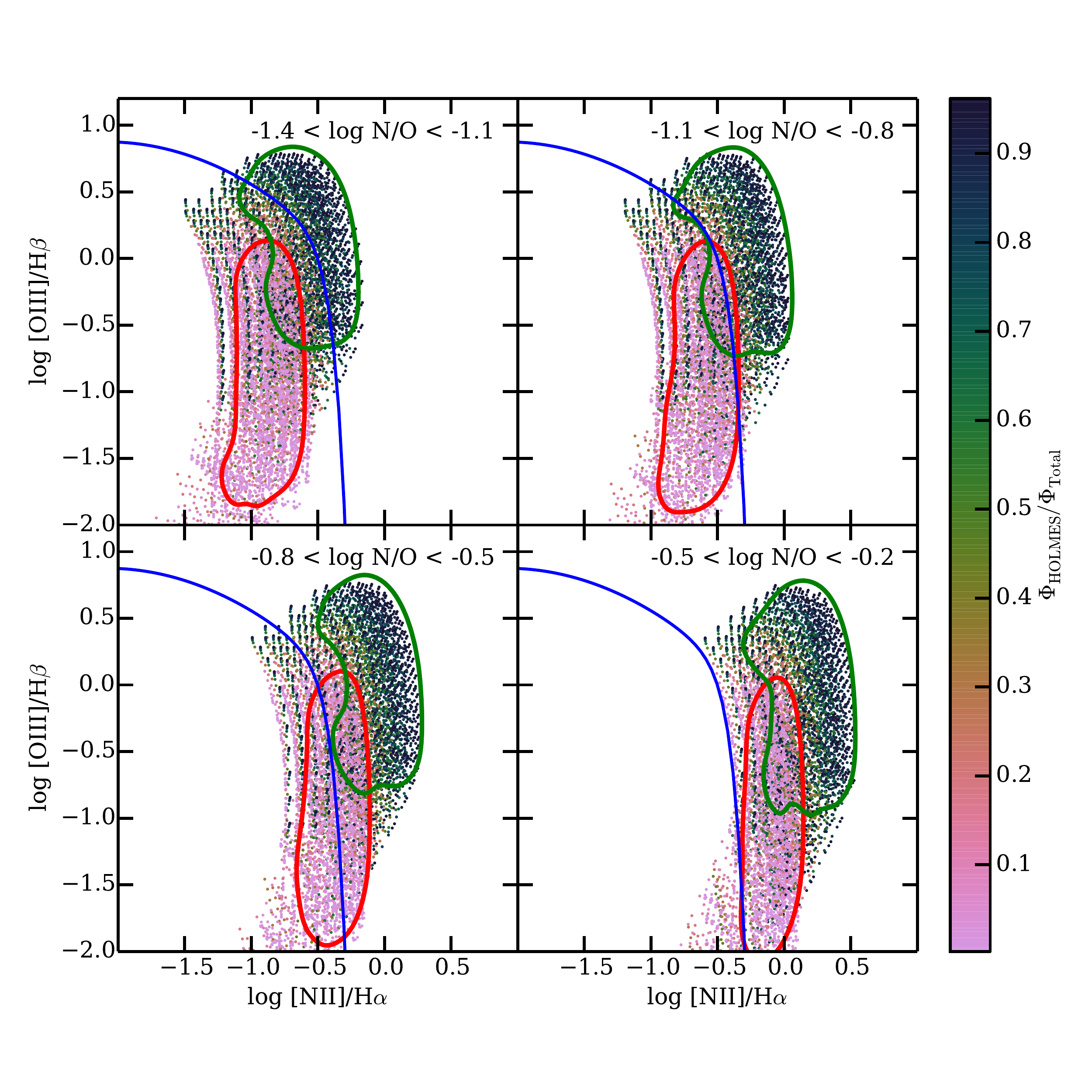}
\caption{[\ion{O}{iii}]/H$\beta$ vs. [\ion{N}{ii}]/H$\beta$  for models referenced as ``DIG\_HR'', for various values of N/O. The blue line correspond to eq.~ 14 of \citet{2006Stasinska_mnra371} dividing star forming galaxies and AGN hosts. Red and green countours have the same definition than in Fig.~\ref{fig:DIG1}.
\label{fig:DIG2} }
\end{figure*}

We use all the ``DIG\_HR'' models (see Sec.\ref{sec:DIGs} and Table \ref{tab:DIG_params}) to illustrate the effect of the hardening of the ionizing radiation on the emission line ratios [\ion{N}{ii}]/H$\alpha$ and [\ion{O}{iii}]/H$\beta$. This hardening occurs when the contribution of the HOLMES to the total ionizing SED increases.
We show two emission-line ratio diagrams \citep[commonly referred to as BPT after][]{1981Baldwin_pasp93}. Fig.~\ref{fig:DIG1} shows [\ion{O}{iii}]/H$\beta$ vs. [\ion{O}{ii}]/H$\beta$ while Fig.~\ref{fig:DIG2} shows [\ion{O}{iii}]/H$\beta$ vs. [\ion{N}{ii}]/H$\alpha$ for different N/O abundance ratios. The colors code the contribution of the HOLMES to the total ionizing flux. The red and green contours contains half of the points for which this contribution is lower than 5\% and greater than 90\% respectively. They are obtained by estimating the probability density function (PDF) using a gaussian kernel density estimation. 

\begin{figure*}\centering
\includegraphics[width=14cm, trim = 0mm 50mm 0mm 10mm]{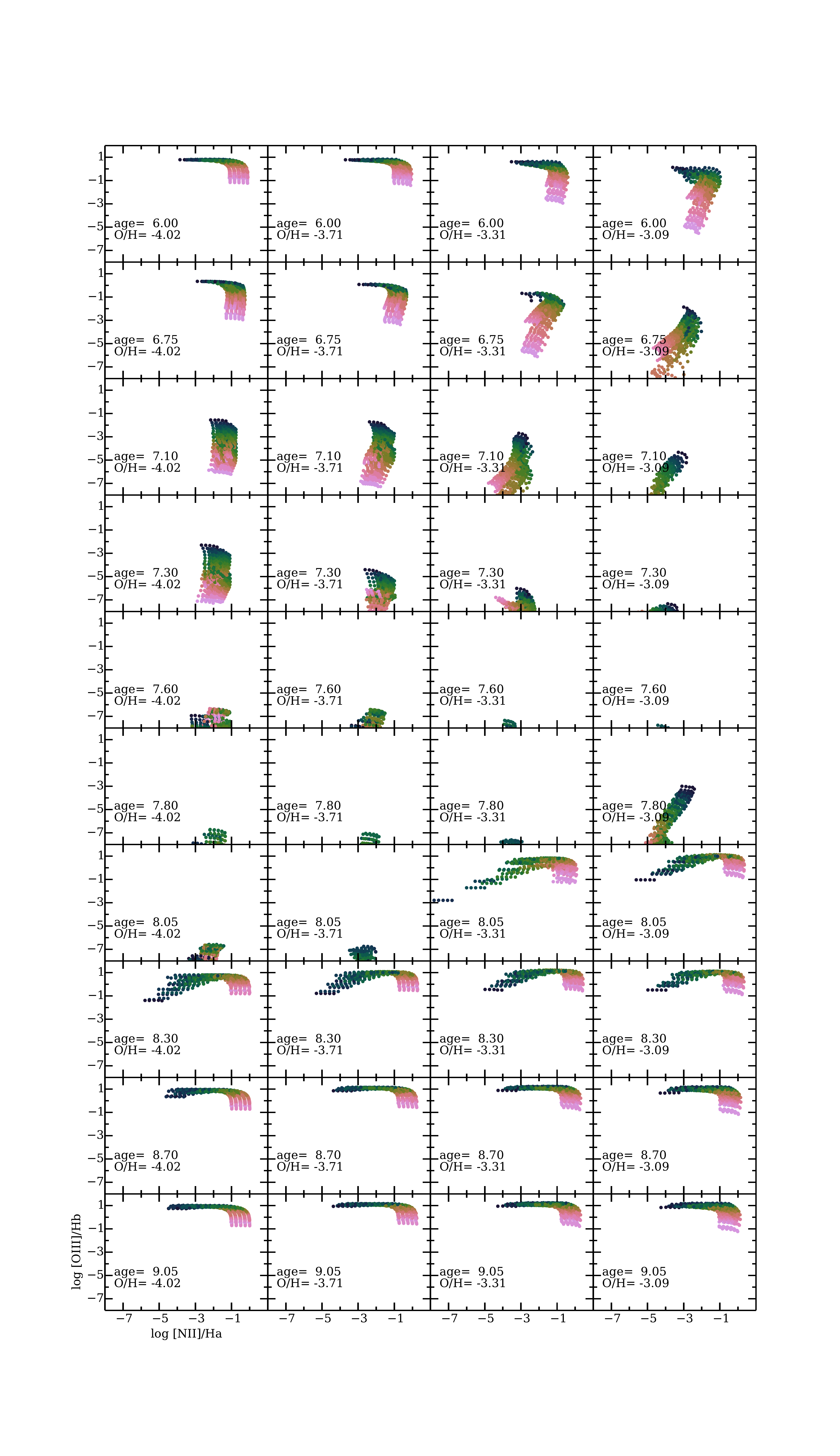}
\caption{BPT diagrams for models referenced as ``CALIFA''. In each panel we use 3 values for N/O, 5 values for the optical depth, 2 values for the geometry and 6 values for the ionization parameter log(U) (this latest being coded by the colors, from light pink corresponding to $-4$ to dark blue corresponding to $-1.5$). Panels from left to right shows the effect of varying the metallicity (of the ionizing SED and the gas) while panels from top to bottom shows the effect of varying the age of the ionizing population. The logarithms of the corresponding values are given in each panel.
\label{fig:CALIFA} }
\end{figure*}

In another example of  the ionization of the interstellar gas by HOLMES, we use the ``CALIFA'' models (see Sec.~\ref{sec:CALIFA}) to plot [\ion{O}{iii}]/H$\beta$ vs. [\ion{N}{ii}]/H$\alpha$ for different ages and metallicities. In each subplot of Fig.~\ref{fig:CALIFA}, the grid is then scanning the effect of varying log(U) and N/O on the two line ratios [\ion{O}{iii}]/H$\beta$ vs. [\ion{N}{ii}]/H$\alpha$. From the left to the right, we see the effect of the metallicity and from top to bottom the effect of the age (only one of three ages is plotted here for concision). 

We see that, for log(age) $\lesssim 6.7$, [\ion{O}{iii}]/H$\beta$ globally decreases, as a consequence of the gradual disappearance of the most massive stars. The effect depends on metallicity. After log(age) $\gtrsim 8$, HOLMES take over the ionization \citep[although at a much reduced pace, not perceptible in emission-line ratio diagrams but see Fig.~2 of][]{2011Cid-Fernandes_mnra413}. The values for [\ion{O}{iii}]/H$\beta$ are higher in this phase than when ionization is due to O-B stars, because of the higher mean energy of the ionizing photons of the HOLMES \citep[see Fig.~4 in][]{2011Flores-Fajardo_mnra415}.
 

\subsection{Combined models of 2 densities}
\label{sec:2comp}

Most of the photoionization models found in the literature are 1D spherical or plane-parallel models. This is actually also the case for the 3MdB models. But if one needs to perform a tailored model to fit detailed and numerous observations of a given object, it often appears that more than one component is needed. It may be necessary to combine models of different optical depth (matter- and radiation-bounded models), models of different densities, models of different chemical compositions, and even any combination using this fundamental differences \citep[e.g.][]{1995Baldwin_apjl455, 1996Binette_aap312, 1996Morisset_aap312, 2002Morisset_aap386, 2003Pequignot_209, 2005Stasinska_aap434, 2010Stasinska_aap511}. 

We can explore the properties of multi component models using models from the 3MdB, in particular we show here models where two regions of different densities are observed together. 
We extract models from the 3MdB under the reference ``PNe\_2014'' (see Sec.~\ref{sec:PNe} and Tab.~\ref{tab:PNe_params}), applying the following criteria: the ionizing source is a blackbody, the models are radiation-bounded, they have constant density and no dust, and they also fit the criteria described in Sec.~\ref{sec:PNe} under the com6=1 condition to represent realistic planetary nebulae. From these models, we select only the ones with lowest hydrogen densities (100, 300 and 1000 cm$^{-3}$), that is 588 models. For each of these models, we look for models corresponding to the same ionizing star and the same chemical composition, but with the highest densities ($3.10^4, 10^5$ and $3.10^5$ cm$^{-3}$). The mean ionization parameter may be different between the 2 regions (depending on the densities, and on where the clump is located with respect to the ionizing star). The total number of models for the clumps under these assumptions is 14620. The models that are obtained by combining the 3MdB models are not included in the 3MdB.

\begin{figure}\centering
\includegraphics[width=8.2cm]{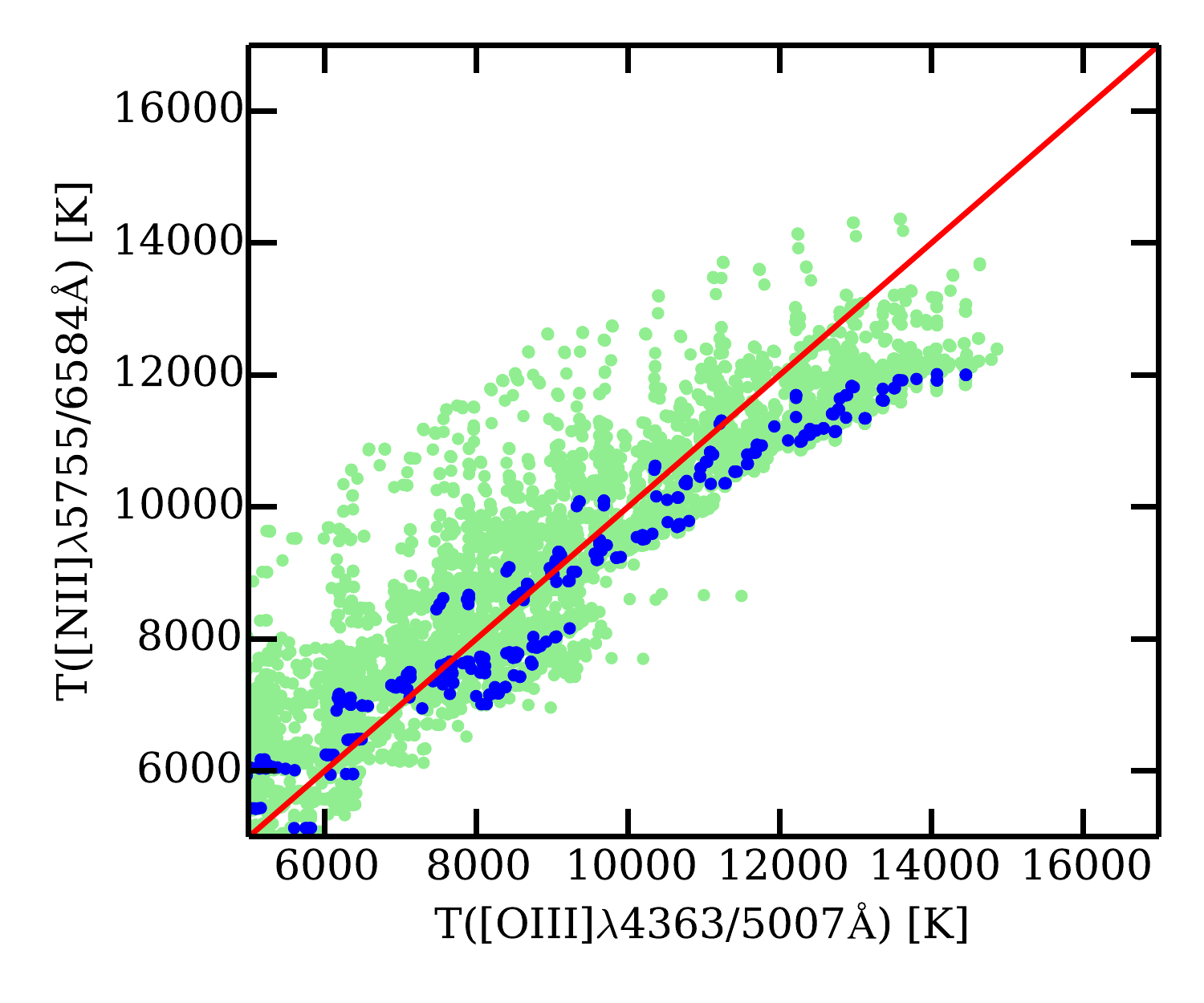}
\includegraphics[width=8.2cm]{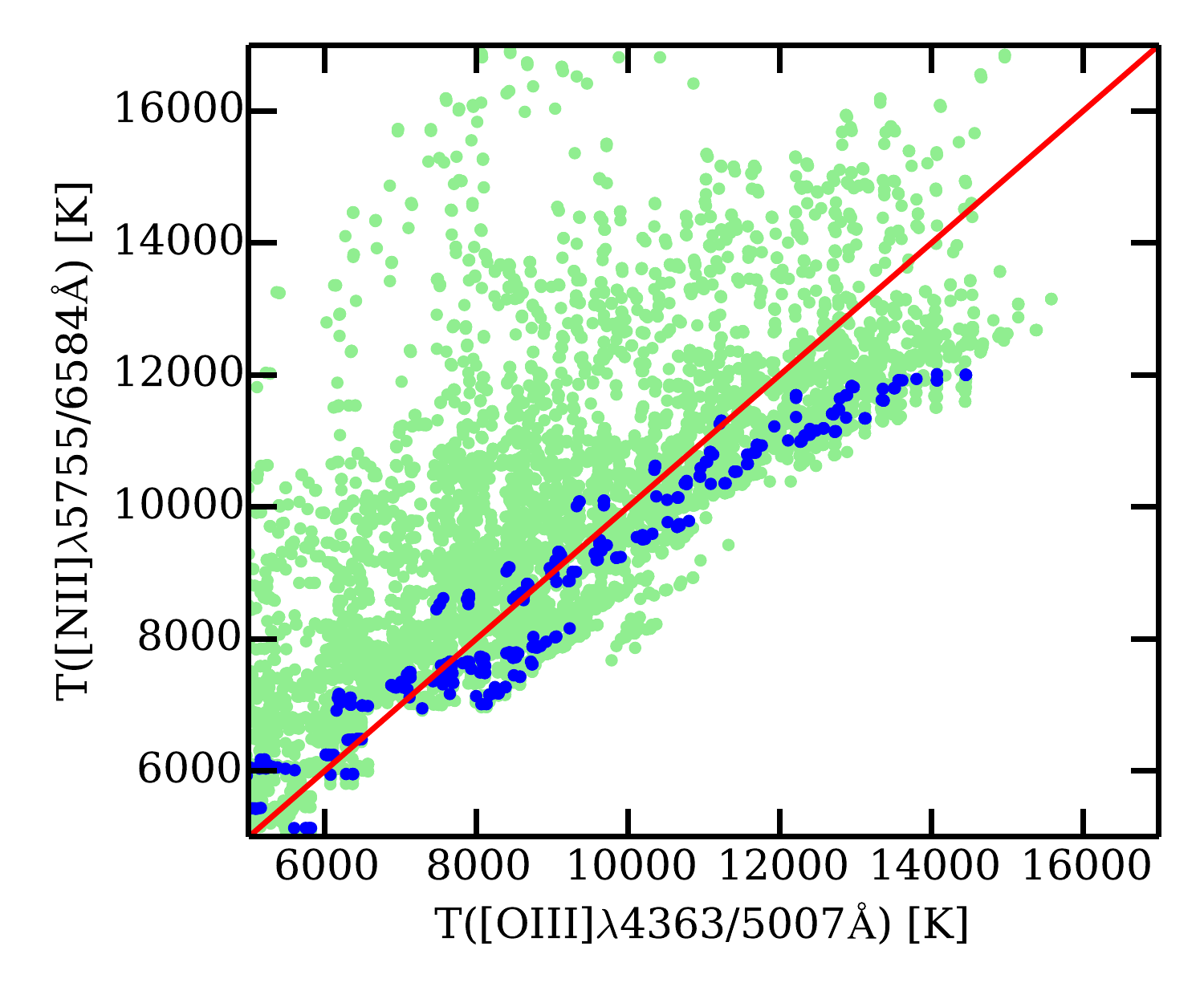}
\caption{Apparent electron temperatures determined by the $[\ion{N}{ii}]\lambda$5755/6584\AA\ and the $[\ion{O}{iii}]\lambda$4363/5007\AA\ line ratios. The dark blue dots are single component low density models, the light green dots are combined models with 5 and 15\% of H$\beta$ comming from dense clumps (upper and lower panel resp.). The red line follows the y=x relation. \label{fig:T2dens} }
\end{figure}

For each pair of models associating a low density and a high density model, we compute two combined models by adding the line emissivities, using a weight for each region so that the contribution of the dense region to the total H$\beta$ intensity is 5  and 15\% respectively. In the following example, we only concentrate on the ``apparent'' electron temperature, as determined from line ratios of classical $[\ion{O}{iii}]\lambda$4363/5007\AA\ and [\ion{N}{ii}]$\lambda$5755/6584\AA\ diagnostics, but any emission line can be computed using the same rule. We use the PyNeb package \citep{2014Luridiana_ArXi} to derive the temperatures from the line ratios of the original and combined models. In Fig.~\ref{fig:T2dens}, we plot the electron temperatures as determined by the [\ion{N}{ii}] line ratio vs. the one obtained from the [\ion{O}{iii}] line ratio. The density used to determine both temperatures is the one obtained from the $[\ion{S}{ii}]\lambda$6731/16\AA\ line ratio. We plot in blue the values obtained for the low density medium alone, and in light green the ones obtained for the combined models. We can see that the values corresponding to the low density medium alone are close to the T([\ion{O}{iii}]) = T([\ion{N}{ii}]) red line. The marginal differences at high temperatures come from the temperature structure of the nebulae, leading to lower temperatures for the outer part of the nebula (the N$^+$ region) compared to the more central part (the O$^{++}$ region). See also Sec.~\ref{sec:TO3N2} on this topic.
The effect of the high density clumps is to artificially increase the apparent [\ion{N}{ii}] temperature. This is due to the fact that the critical density of the level 4 of N$^+$ is $\sim 10^5 $cm$^{-3}$, while the corresponding critical density for the O$^{++}$ ion is  $\sim 7.10^5 $cm$^{-3}$ (level 5 critical densities being of order of $10^7 $cm$^{-3}$). The $[\ion{N}{ii}]\lambda$6584\AA\ emissivity is thus strongly reduced relative to the $[\ion{N}{ii}]\lambda$5755\AA\ one in clumps of density of order of $10^5 $cm$^{-3}$, leading to a higher value of T([\ion{N}{ii}]).

High T([\ion{N}{ii}])/T([\ion{O}{iii}]) are observed in high excitation planetary nebulae with Wolf-Rayet central stars \citep{2001Pena_aap367, 2012Garcia-Rojas_aap538} and no satisfying explanation was given for that so far. The inclusion of dense clumps such as illustrated above might be the answer.  

\subsection{Temperatures of the N$^+$ and the O$^{++}$ regions  in the different projects}
\label{sec:TO3N2}

\begin{figure*}\centering
\includegraphics[width=16cm]{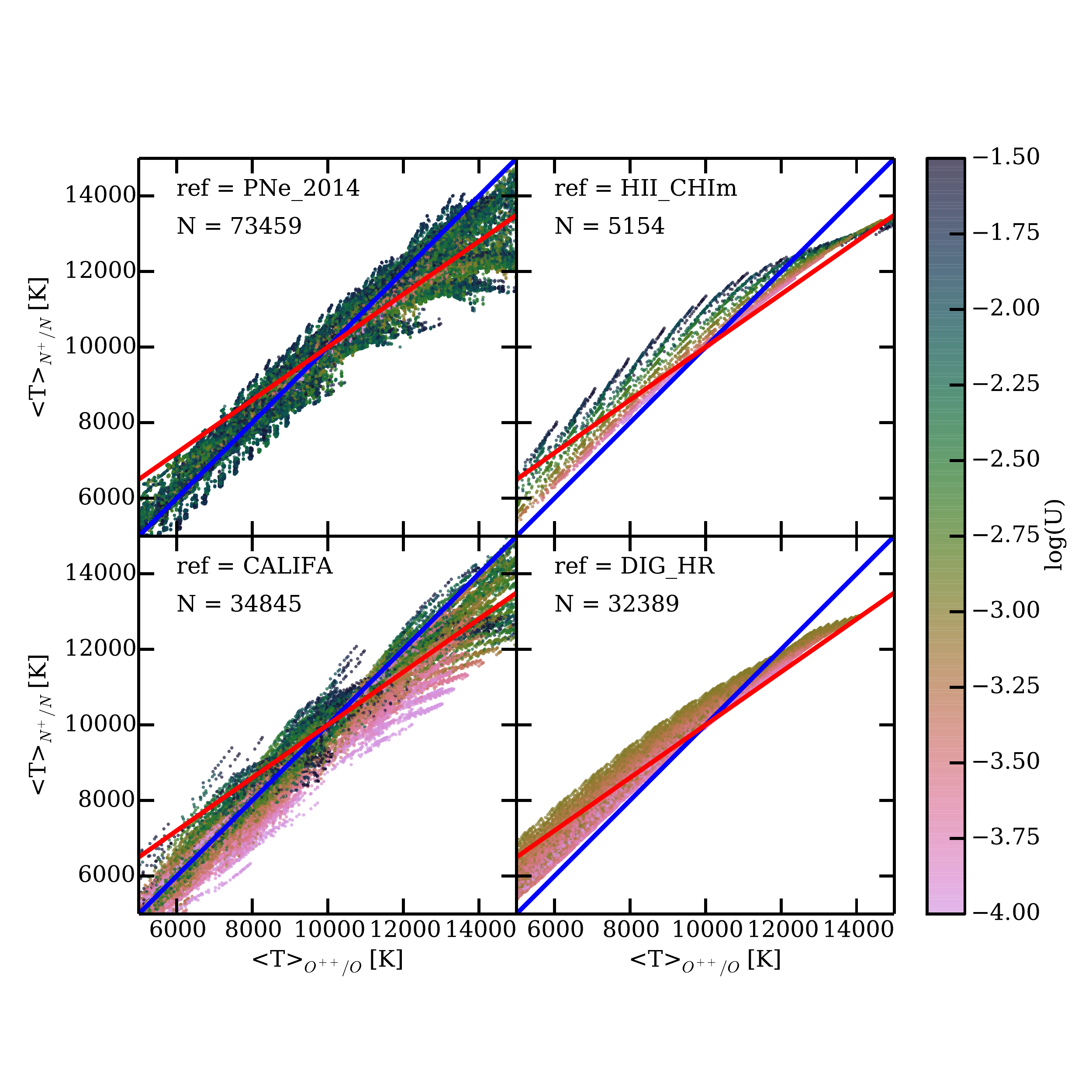}
\caption{Mean electron temperature weighted by the O$^{++}$/O ionic fraction versus weighted by N$^{+}$/N, for the four projects held in the 3MdB. The colors code the ionization parameter log(U) in the same way in every panel. The blue line shows the equal temperatures location. In the upper-left part of each panel the reference of the project and the number of points plotted are given. See text for more details on the filters applied to the models before plotting them. The blue line follows y=x, while the red line is following the relation $y = 0.7 \times x + 3000$ from \citet{1986Campbell_mnra223} .
\label{fig:TO3N2}}
\end{figure*}

One of the major advantage of the 3MdB is that all the models of the different projects are in a universal environment, allowing easy comparison. In this example, we plot in Fig.~\ref{fig:TO3N2} the mean electron temperature weighted by the O$^{++}$/O ionic fraction versus the one weighted by N$^{+}$/N. The plots are done separately for the four projects currently in the 3MdB. In all cases, we filter the models to only plot the ones that have $[\ion{O}{iii}]\lambda$5007/H$\beta > 0.05$ and $[\ion{N}{ii}]\lambda$6584/H$\beta > 0.05$, to avoid models where the O$^{++}$ or the N$^{+}$ regions would be too small. In the case of the ``PNe\_2014'' models, we filter the models with the com6 = 1 condition (the realistic models, see Sec.\ref{sec:PNe}). The blue line follows y=x, while the red line is following the relation $y = 0.7 \times x + 3000$ from \citet{1986Campbell_mnra223}.
Some trends are clearly seen:
\begin{itemize}
\item All the models are showing the expected trend of both temperatures increasing together.
\item Neither of the two lines is a good fit to the points from the models, a somewhat large scatter is observed.
\item In the case of ``HII\_CHIm'' models, only one SED has been used and the departure from the y=x line is increasing with the ionization parameter. 
\item In the case of ``DIG\_HR'' models, the ionization parameter is only scanning one order of magnitude (from $-4$ to $-3$), and the departure from the y=x is mainly due to the shape of the SED, changing from dominated by OB stars to dominated by HOLMES (not shown in this plots).
\item In the cases of ``PNe\_2014'' and ``CALIFA'' models, the SED and log(U) are changed and the result is a larger scatter around the y=x line, without a clear effect of log(U) alone.
\item In all the cases,  we do not see very high values of ${<T>}_{N^{+}/N} / {<T>}_{O^{++}/O}$ at ${<T>}_{O^{++}/O}$ larger than 10,000~K. The observed high values for this ratio in some planetary nebulae in this temperature range may be explained by the presence of high density clumps, see Sec.~\ref{sec:2comp}.
\end{itemize}

As an example of the use of MySQL to request data from the 3MdB, the following is the request for the lowest-right panel:
\begin{verbatim}
SELECT T_NITROGEN_vol_1, T_OXYGEN_vol_2, 

logU_mean, O__3__5007A, TOTL__4861A, 

N__2__6584A FROM tab, teion WHERE 

(tab.ref = 'HII_CHIm' and 

O__3__5007A/H__1__4861A > 0.05 and 

N__2__6584A/H__1__4861A > 0.05 and 

tab.N = teion.N );
\end{verbatim}

Notice the use of the field ``N'' to join the tables ``tab'' and ``teion''. In the case of the ``PNe\_2014'' models, a condition {\rm com6 = 1} has been added. The request results weight 6.1Mo, 441Ko, 957Ko, and 2.7Mo for ref = ``PNe\_2014'', ``HII\_CHIm'', ``CALIFA'', and ``DIG\_HR'' respectively.

\section{Conclusions}

We have presented in this paper the 3MdB, a new tool in the form of a database, to deal with grids of photoionization models. This tool is accessible with the MySQL protocol through Internet. It holds hundreds of thousands of photoionization models computed with the {\sc CLOUDY} program \citep{2013Ferland_rmxa49}. For each model, more than 2600 parameters and model outputs are available. The power of the MySQL system allows the user to make complex requests based on filters to obtain subsets of this huge amount of data, in a format easy to read (e. g. coma-separated-variable ascii format). The 3MdB system currently holds 4 different projects (sets of data), other projects are planned to be incorporated in a near future. Up-to-date information is available from the 3MdB web page (\url{https://sites.google.com/site/mexicanmillionmodels/}). 

We have shown with a few examples that the database can provide a very user-friendly way to study some aspects of the physics of the interstellar medium. Those examples are only illustrative, some of them will be developed in forthcoming papers. 

It is worth noticing that the user is {\it in fine} responsible of the coherence of the work he is doing with the data extracted from the 3MdB. In particular, some care must be taken when comparing models computed with different versions of {\sc CLOUDY}, or when dealing with models from the border of a grid, where the set of parameters may define irrealistic nebulae. In any case, for each particular project, the user should carefully check the relevance of the parameter space explored with the 3MdB subgrid used. The user needs to keep in mind that the models held in 3MdB suffer the limitations inherent to any photoionization model made with {\sc CLOUDY} or with any other code: uncertainties in the atomic data or in the ionizing spectral energy distribution obtained from atmosphere models, optical properties of dust, among others.

The 3MdB is of public access through the MySQL protocol. The data are under the BSD-new license. For security reasons, the server IP, username and password for the direct access to the MySQL server are now only provided after asking by email to the authors. A user-web interface will be given as soon as possible, with full open access to the database.

One of the future extensions of the 3MdB is to be included as a Virtual Observatory service. This will provide more interfaces for the user to access the data.

Colleagues interested to include their grids of photoionization models in the 3MdB (because they do not have the computational ressources to run a large amount of models, to ease the accessibility to their results, or to insure the sustainability of their grid) are welcome to contact CM.

\section{Acknowledgments}
CM, GDI and NFF acknowledge support from the mexican project CONACyT CB-2010/153985. GDI gratefully aknowledges the DGAPA postdoctoral grant from the Universidad Nacional Autónoma de México. Many thanks to Gra\.zyna Stasi\'nska who kindly read and improved the manuscript. The authors gratefully acknowledge the constructive comments from the referee.

\bibliography{3mdb}

\end{document}